\documentclass[preprint]{aastex}

\shorttitle{High D$_2$O/HDO ratio in the inner regions of the low-mass protostar NGC1333 IRAS2A}
\shortauthors{Coutens et al.}

\begin{document}


\title{High D$_2$O/HDO ratio in the inner regions  \\
of the low-mass protostar NGC1333 IRAS2A\thanks{Based on observations carried out with the IRAM Plateau de Bure Interferometer. IRAM is supported by INSU/CNRS (France), MPG (Germany) and IGN (Spain).}}


\author{A. Coutens\altaffilmark{1,2}, J. K. J\o rgensen\altaffilmark{1,2}, M. V. Persson\altaffilmark{3}, E. F. van Dishoeck\altaffilmark{3,4}, C. Vastel\altaffilmark{5,6}, V. Taquet\altaffilmark{7}}
\email{acoutens@nbi.dk}
\altaffiltext{1}{Niels Bohr Institute, University of Copenhagen, Juliane Maries Vej 30, DK-2100 Copenhagen \O, Denmark  }
\altaffiltext{2}{Centre for Star and Planet Formation, Natural History Museum of Denmark, University of Copenhagen, \O ster Voldgade 5-7, DK-1350 Copenhagen K, Denmark }
\altaffiltext{3}{Leiden Observatory, Leiden University, PO Box 9513, 2300 RA Leiden, The Netherlands }
\altaffiltext{4}{Max-Planck-Institut f\"ur Extraterrestrische Physik, Giessenbachstrasse 1, 85748 Garching, Germany }
\altaffiltext{5}{Universit\'e de Toulouse, UPS-OMP, IRAP, Toulouse, France}
\altaffiltext{6}{CNRS, IRAP, 9 Av. Colonel Roche, BP 44346, 31028 Toulouse Cedex 4, France}
\altaffiltext{7}{NASA Postdoctoral Program Fellow, NASA Goddard Space Flight Center, 8800 Greenbelt Road, Greenbelt, MD 20770, USA}


\begin{abstract}
Water plays a crucial role both in the interstellar medium and on Earth. To constrain its formation mechanisms and its evolution through the star formation process, the determination of the water deuterium fractionation ratios is particularly suitable. Previous studies derived HDO/H$_2$O ratios in the warm inner regions of low-mass protostars. 
We here report a detection of the D$_2$O 1$_{1,0}$--1$_{0,1}$ transition toward the low-mass protostar NGC1333~IRAS2A with the Plateau de Bure interferometer: this represents the first interferometric detection of D$_2$O -- and only the second solar-type protostar for which this isotopologue is detected.
Using the observations of the HDO 5$_{4,2}$--6$_{3,3}$ transition simultaneously detected and three other HDO lines previously observed, we show that the HDO line fluxes are well reproduced with a single excitation temperature of 218$\pm$21\,K and a source size of $\sim$0.5$\arcsec$. The D$_2$O/HDO ratio is $\sim$(1.2$\pm$0.5)\,$\times$\,10$^{-2}$, while the use of previous H$_2^{18}$O observations give an HDO/H$_2$O ratio of $\sim$(1.7$\pm$0.8)\,$\times$\,10$^{-3}$, i.e. a factor of 7 lower than the D$_2$O/HDO ratio. These results contradict the predictions of current grain surface chemical models and indicate that either the surface deuteration processes are poorly understood or that both sublimation of grain mantles and water formation at high temperatures ($\gtrsim$230\,K) take place in the inner regions of this source. In the second scenario, the thermal desorption of the grain mantles would explain the high D$_2$O/HDO ratio, while water formation at high temperature would explain significant extra production of H$_2$O leading to a decrease of the HDO/H$_2$O ratio.
\end{abstract}


\section{Introduction}

In addition to being a crucial ingredient for the emergence of life on Earth, water is omnipresent in the interstellar medium and is detected both in the gas and solid phases (e.g., \citealt{vanDishoeck2013}). Observations of its deuterated forms are necessary to understand how water forms and how it evolves during star formation until its incorporation into comets and asteroids that may then deliver water to planets through impacts \citep[e.g.,][]{Raymond2004,Hartogh2011}. Indeed, the water deuterium fractionation is quite sensitive to the conditions (temperature and density) in which water forms.
The relative abundance ratios between H$_2$O, HDO, and D$_2$O act then as a tracer of the water formation and alteration (e.g., \citealt{Ceccarelli2014}). They can be estimated at different stages of the star formation process in order to constrain the water evolution history. 

Numerous lines of HDO have been detected towards low-mass protostars with ground-based single-dish telescopes \citep{Stark2004,Parise2005,Liu2011} as well as the Heterodyne Instrument for Far-Infrared (HIFI)  onboard the {\it Herschel} Space Observatory \citep{Coutens2012,Coutens2013b}.
A few excited lines have also been detected with interferometers \citep{Codella2010,Persson2013,Persson2014,Taquet2013}, showing that their emission mainly arises from the warm inner regions of low-mass protostars.

Due to its double deuteration, heavy water (D$_2$O) is more challenging to detect. The para--D$_2$O 1$_{1,0}$--1$_{0,1}$ fundamental transition was, however, detected at 316.8 GHz with the \textit{James Clerck Maxwell} Telescope  (JCMT) towards the Class~0 protostar IRAS~16293-2422 by \citet{Butner2007}.
Two additional fundamental transitions (1$_{1,1}$--0$_{0,0}$ and 2$_{1,2}$--1$_{0,1}$) were detected in absorption towards the same source with the HIFI instrument at 607.3 and 897.9 GHz, respectively \citep{Vastel2010,Coutens2013a}. These absorption components, also seen for HDO, arise from a cold extended layer surrounding the protostar, where gaseous water is assumed to be produced by photodesorption mechanisms (cosmic-ray induced and/or external UV field). Thanks to 1D radiative transfer modeling of these three D$_2$O lines and other HDO and H$_2^{18}$O lines \citep{Coutens2012,Coutens2013a}, the water deuterium fractionation was found to be extremely high in this cold layer (HDO/H$_2$O\,$\sim$\,5\% and D$_2$O/HDO\,$\sim$\,11\%). It seems, however, to decrease in the warm inner regions, as suggested by the upper limit of 0.9\% derived for D$_2$O/HDO. The complete ice mantle sublimates at dust temperatures higher than 100 K \citep{Fraser2001}. The D/H ratios derived in these warm inner regions should then reflect the overall (gas + grain) deuteration of water.
To constrain the water formation mechanisms in the inner protostellar region, it is, however, required to determine more precisely the D$_2$O/HDO ratio. 

We report, in this letter, the first detection of D$_2$O towards the Class 0 protostar NGC1333~IRAS2A. This source is only the second low-mass protostar in which D$_2$O is detected. It is also the first interferometric observation of D$_2$O. 
The HDO 5$_{4,2}$--6$_{3,3}$ transition at 317.2 GHz was also detected for the first time and is, to date, the most highly excited transition of HDO ($E_{\rm up}$\,=\,691\,K) ever detected in a low-mass protostar. From these observations, we determine the D$_2$O/HDO ratio in the warm inner region of NGC1333~IRAS2A, providing important constraints on water formation.

\section{Observations}

Observations of the low-mass protostar NGC1333 IRAS2A were carried out with the IRAM Plateau de Bure Interferometer (PdBI) on 2013 December 3$^{\rm rd}$ in the C configuration under very good weather conditions. 
The para--D$_2$O 1$_{1,0}$--1$_{0,1}$ transition was observed at 316.8 GHz using the narrow-band correlator with a spectral resolution of 0.078 MHz ($dv$ = 0.07 km\,s$^{-1}$). 
The observations of the continuum emission and the HDO 5$_{4,2}$--6$_{3,3}$ transition at 317.2 GHz have been obtained simultaneously using the WIDEX correlator that provides a spectral resolution of 1.95 MHz ($dv$ = 1.8 km\,s$^{-1}$ at 317.2 GHz).
The data were calibrated and imaged using the CLIC and MAPPING packages of the GILDAS\footnote{\url{http://www.iram.fr/IRAMFR/GILDAS/}} software. 
Phase and amplitude were calibrated by observing the nearby strong quasars 0333+321 and 3C84.
The bandpass calibration was carried out on 3C454.3, and the absolute flux density scale was derived from LkH$\alpha$~101 and MWC 349.
The continuum was subtracted before Fourier transformation of the line data.

\begin{table}[t]
\caption{Parameters from an elliptical Gaussian fit to the Continuum Emission.} 
\label{emission}
\begin{center}
\begin{tabular}{l c}
\tableline\tableline
& IRAS2A \\
\tableline
Synthesized beam & 0.89$\arcsec$ $\times$ 0.76$\arcsec$ (+42$^\circ$) \\
 Flux (Jy)\tablenotemark{a} & 0.578 ($\pm$ 0.003)\\
 RA (J2000)\tablenotemark{a} & 03:28:55.57 ($\pm$ 0$\arcsec$.002) \\
 Dec (J2000)\tablenotemark{a} & 31:14:37.07 ($\pm$ 0$\arcsec$.002) \\
 Extent\tablenotemark{b} & 0.95$\arcsec$ $\times$ 0.68$\arcsec$ (+38$^\circ$) \\
\tableline 
\end{tabular}
\tablenotetext{a}{ Errors given are the statistical errors from the fitting routine.} 
\tablenotetext{b}{Major $\times$ Minor axis and position angle for elliptical Gaussian \\ profile 
in the ($u,v$)-plane.}
\end{center}
\end{table}

\begin{figure}[t!]
\includegraphics[angle=0,scale=.5]{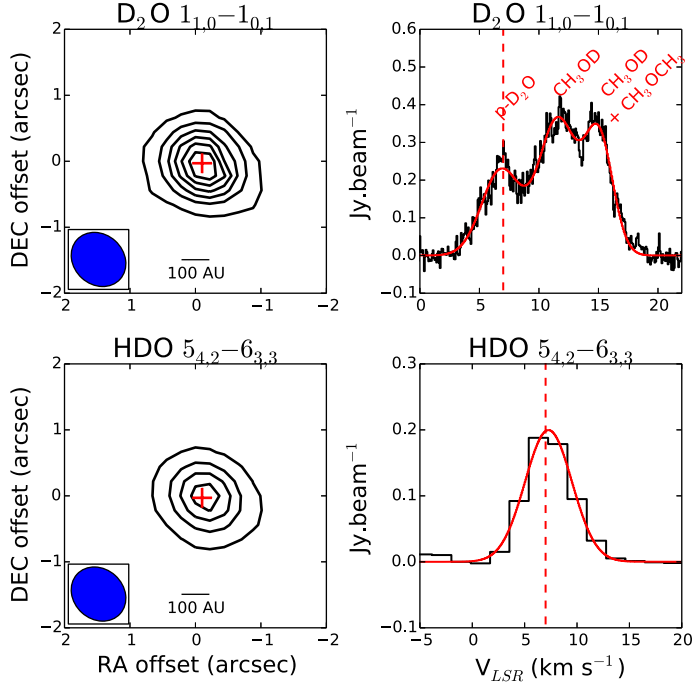}
\caption{\textit{Left:} Integrated intensity maps for the para--D$_2$O 1$_{1,0}$--1$_{0,1}$ and HDO 5$_{4,2}$--6$_{3,3}$ lines. The reference position (0,0) is $\alpha_{2000}$=03$^{\rm h}$28$^{\rm m}$55$\fs$58, $\delta_{2000}$=+31\degr$14\arcmin$37.1$\arcsec$. The
beam is shown in the lower left corner. The red cross shows the position of the peak of the continuum. Contours start at 50 mJy\,km\,s$^{-1}$ (5$\sigma$) with steps of 150 mJy\,km\,s$^{-1}$. The integrated intensity map of D$_2$O is obtained using the three-Gaussian decomposition of the line profile. \textit{Right:} Continuum subtracted spectra of the para--D$_2$O 1$_{1,0}$--1$_{0,1}$ and HDO 5$_{4,2}$--6$_{3,3}$ lines (in black). The red line shows the results of the Gaussian fit. }
\label{figure_obs}
\end{figure}

The parameters of the continuum emission derived with an elliptical Gaussian fit to the ($u,v$)-plane are presented in Table \ref{emission}.
The D$_2$O line is detected at the expected velocity of the source ($v_{LSR}$\,=\,7.0\,km\,s$^{-1}$). It is, however, blended with two other components at a velocity of $\sim$11.5\,km\,s$^{-1}$ and $\sim$15.0\,km\,s$^{-1}$ (see Figure \ref{figure_obs}). The first component is identified as the transition 7$_0$--6$_0$+ of CH$_3$OD ($\nu$\,=\,316.7951\,GHz, \citealt{Anderson1993}).
 The second component at 15\,km\,s$^{-1}$ corresponds to the CH$_3$OD 3$_2$--3$_1$+ transition ($\nu$\,=\,316.7916\,GHz), but the CH$_3$OCH$_3$ 22$_{6,16}$--22$_{5,17}$ transitions ($\nu$\,=\,316.7908, 316.7918, 316.7928\,GHz) can also contribute to the line flux of this component.  
 The presence of other species than D$_2$O around 7\,km\,s$^{-1}$ can be excluded using the spectroscopic databases. Although some CH$_3$CHO lines are detected in the WIDEX data, a local thermal equilibrium (LTE) modeling of these lines does not predict any emission for the 9$_{1,8,8}$--8$_{1,8,7}$ transition ($\nu$\,=\,316.8034\,GHz) at a velocity of 7.9\,km\,s$^{-1}$. The SO$^{17}$O 11$_{2,10}$--10$_{1,9}$ transition ($\nu$\,=\,316.7988\,GHz) cannot contribute to the line flux at 9.5\,km\,s$^{-1}$, as the equivalent transition of the more abundant isotopologue $^{34}$SO$_2$ is not detected in the WIDEX data.

\begin{table*}[t!]
\caption{Parameters of the D$_2$O, HDO, and H$_2^{18}$O observed lines.$^1$}
\label{table_parameters}
\begin{center}
\begin{tabular}{l c r c c c c c c }
\tableline\tableline
Transition & Frequency & $E_{\rm up}$ & $A_{\rm ij}$ & Beam  & Flux$^2$ & Size$^3$ & FWHM$^4$ & Ref.$^5$ \\
& (GHz) & (K) & (s$^{-1}$) & ($\arcsec$ $\times$ $\arcsec$) &  (Jy\,km\,s$^{-1}$) & ($\arcsec$) & (km\,s$^{-1}$) &   \\
\hline
p--D$_2$O 1$_{1,0}$--1$_{0,1}$ & 316.7998 & 15 & 6.4\,$\times$\,10$^{-4}$ & 0.9\,$\times$\,0.8 & 0.74\,$\pm$\,0.15 & ...$^6$ & 3.8\,$\pm$\,0.2  &  a\\
\hline
HDO 5$_{4,2}$--6$_{3,3}$ & 317.1512 & 691 & 2.7\,$\times$\,10$^{-5}$ & 0.9\,$\times$\,0.8 & 1.10\,$\pm$\,0.22 & ...$^6$ & 5.3\,$\pm$\,0.2 & a\\
HDO 4$_{2,2}$--4$_{2,3}$ & 143.7272 & 319 & 2.8\,$\times$\,10$^{-6}$ & 2.2\,$\times$1.8 & 0.67\,$\pm$\,0.13  & 0.6\,$\pm$\,0.2 & 6.7\,$\pm$\,0.5 & b\\
HDO 3$_{1,2}$--2$_{2,1}$ & 225.8967& 168 & 1.3\,$\times$\,10$^{-5}$ & 1.3\,$\times$\,1.0 & 3.98\,$\pm$\,0.80 & 0.4\,$\pm$\,0.1 & 4.1\,$\pm$\,0.1 & c\\
HDO 2$_{1,1}$--2$_{1,2}$ & 241.5616 & 95  & 1.2\,$\times$\,10$^{-5}$ & 1.2\,$\times$\,0.9 & 3.88\,$\pm$\,0.78 & 0.5\,$\pm$\,0.1 & 4.0\,$\pm$\,0.1 & c\\
\hline
p--H$_2^{18}$O 3$_{1,3}$--2$_{2,0}$ & 203.4075 & 204 & 4.8\,$\times$\,10$^{-6}$ & 0.9\,$\times$\,0.7 & 0.98\,$\pm$\,0.20 & 0.8\,$\pm$\,0.1 & 4.0\,$\pm$\,0.1 & d\\
\tableline 
\end{tabular}
\tablenotetext{1}{The HDO, D$_2$O, and H$_2^{18}$O spectroscopic parameters come from the JPL and CDMS databases \citep{Pickett1998,Muller2005}.}
\tablenotetext{2}{An observational uncertainty of 20\% is assumed for each line.}
\tablenotetext{3}{FWHM of circular Gaussian fit in the $(u,v)$-plane. The data used in \citet{Taquet2013} were reprocessed and the line flux re-measured.}
\tablenotetext{4}{FWHM measured towards the pixel showing the peak in emission. The higher FWHM for the HDO lines at 143.7 and 317.2 GHz are due to the lower spectral resolution of the WIDEX data (1.95\,MHz).}
\tablenotetext{5}{References: a) This study; b) \citet{Taquet2013}; c) \citet{Persson2014}; d) \citet{Persson2012}}
\tablenotetext{6}{The $(u,v)$-plane is best fitted by a point source.}
\end{center}
\end{table*}

The total flux of the para--D$_2$O line was extracted with a three-Gaussian decomposition on each pixel where emission is detected, and is estimated to be 0.74\,Jy\,km\,s$^{-1}$. The flux of the HDO line was measured in the ($u,v$)-plane and is about 1.1\,Jy\,km\,s$^{-1}$. A flux calibration uncertainty of 20\% is assumed.
The D$_2$O and HDO emission peaks are situated at the same position as the peak of the continuum emission (see Figure \ref{figure_obs}).

Three other HDO transitions were previously detected with the PdBI: the HDO 3$_{1,2}$--2$_{2,1}$ and 2$_{1,1}$--2$_{1,2}$ transitions at 225.9 and 241.6 GHz, respectively, by \citet{Persson2014} and the 4$_{2,2}$--4$_{2,3}$ transition at 143.7 GHz by \citet{Taquet2013}. The H$_2^{18}$O 3$_{1,3}$--2$_{2,0}$ line was also detected at 203.4 GHz by \citet{Persson2012}. The parameters of the different lines are summarized in Table \ref{table_parameters}.
The different water isotopologues show full widths at half maximum (FWHM) of $\sim$4\,km\,s$^{-1}$. The HDO lines at 317.2 and 143.7 GHz only show higher widths because of the lower spectral resolution of the WIDEX data.

\section{Results}

\begin{figure}[t!]
\includegraphics[angle=0,scale=.4]{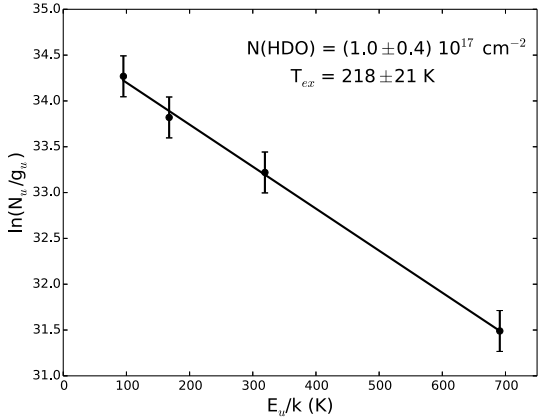}
\caption{Rotational diagram of HDO for a source size of 0.5$\arcsec$.}
\label{figure_RD_hdo}
\end{figure}

Using the four HDO detections, we carried out a LTE analysis through the rotational diagram method \citep{Goldsmith1999}. Because of the different beams of the observations, we considered beam dilution factors for different source sizes. We determined the best-fit through a $\chi^2$ minimization. Good linear fits are obtained for source sizes smaller than 1$\arcsec$. The best agreement is found for a source size of 0.5$\arcsec$ (see Figure \ref{figure_RD_hdo}). It is relatively close to the extent (0.4$\arcsec$--0.6$\arcsec$) determined by a Gaussian fit in the ($u,v$)-plane for the HDO lines at 225.9, 241.6 and 143.7 GHz. 
The column densities and excitation temperature derived are summarized in Table \ref{table_RD}. 
For source sizes smaller than 0.3$\arcsec$, the optical thickness of the HDO lines at 225.9 and 241.6 GHz become non-negligible, and the four HDO line fluxes cannot be reproduced simultaneously anymore with a single excitation temperature. The similarity of the fitted line widths of the observed transitions is also favorable to their optical thinness or, at most, moderate optical thickness (see Table \ref{table_parameters}).

Using the flux derived from our D$_2$O observations and the same excitation temperature as HDO, we calculated the para--D$_2$O column density as a function of the source size (see Table \ref{table_RD}). Assuming that the ortho/para(D$_2$O) ratio is 2:1 (LTE value at T\,$\gtrsim$\,20\,K), we determined a D$_2$O/HDO ratio of (1.2\,$\pm$\,0.5)\,$\times$\,10$^{-2}$. 
We used the same procedure to estimate the para--H$_2^{18}$O column density, using the flux of the transition 3$_{1,3}$--2$_{2,0}$ detected by \citet{Persson2012}.
Assuming that the ortho/para ratio of H$_2^{18}$O is 3:1, i.e. the equilibrium value at high temperature (T\,$>$\,50\,K), and that H$_2^{16}$O/H$_2^{18}$O is 560 \citep{Wilson1999}, the HDO/H$_2$O ratio is about (1.7\,$\pm$\,0.8)\,$\times$\,10$^{-3}$. Neither the D$_2$O line nor the H$_2^{18}$O line are optically thick for the considered source sizes.

As the upper energy level of the D$_2$O transition is lower
  (15\,K) than those of the other lines (95--691\,K), we also checked
  that a lower excitation temperature would not decrease considerably
  the D$_2$O/HDO ratio. To get a D$_2$O/HDO ratio as low as that of
  HDO/H$_2$O, the D$_2$O excitation temperature should be lowered to
  30\,K. This is excluded in the warm inner regions of protostars
  ($T$\,$>$\,100\,K) and could only occur if most of the D$_2$O
  emission arises from the cold outer envelope, which seems inconsistent
  with the observed spatial distribution.

With the non-LTE RADEX code \citep{vanderTak2007}, the fluxes of the HDO lines at 225.9, 241.6 and 143.7 GHz can only be reproduced when the H$_2$ density and the temperature are higher than 2\,$\times$\,10$^{8}$\,cm$^{-3}$ and  125\,K, respectively. The flux of the HDO line at 317.2 GHz cannot be estimated with this method, as the collisional coefficients (with ortho and para--H$_2$) are only determined for the transitions with $E_{\rm up}$\,$\leq$\,444\,K \citep{Faure2012}.
The derived HDO and H$_2^{18}$O column densities are similar to the LTE values presented earlier. The D$_2$O column density cannot, however, be determined with a non-LTE code, as the collisional coefficients available for D$_2$O \citep{Faure2012} are only calculated with para--H$_2$ for the first six energy levels ($E_{\rm up}$\,$<$\,112\,K). They are thus not suited for estimates in warm regions ($T$\,$\gtrsim$\,100\,K).

\begin{table*}[t!]
\caption{Water deuterium fractionation derived with the LTE analysis.$^1$}
\label{table_RD}
\begin{center}
\begin{tabular}{c | c c c | c | c | c c c c }
\tableline\tableline
Source  & N(HDO) & T$_{\rm ex}$ & $\chi^2$ & N(p--D$_2$O) & N(p--H$_2^{18}$O) & D$_2$O/HDO & HDO/H$_2$O & D$_2$O/H$_2$O \\
size ($\arcsec$) & ($\times$10$^{17}$\,cm$^{-2}$) & (K) & & ($\times$10$^{14}$\,cm$^{-2}$) & ($\times$10$^{16}$\,cm$^{-2}$) & ($\times$10$^{-2}$) & ($\times$10$^{-3}$) & ($\times$10$^{-5}$)\\
\hline
0.3~ & 2.5\,$\pm$\,0.9 & 212\,$\pm$\,20 & 0.45 & 9.4\,$\pm$\,2.0 & 6.1\,$\pm$\,1.6 & 1.1\,$\pm$\,0.5 & 1.8\,$\pm$\,0.8 & 2.0\,$\pm$\,0.7\\
0.4~ & 1.5\,$\pm$\,0.5 & 215\,$\pm$\,20 & 0.28 & 5.9\,$\pm$\,1.2 & 3.8\,$\pm$\,1.1 & 1.2\,$\pm$\,0.5 & 1.8\,$\pm$\,0.8 & 2.1\,$\pm$\,0.7 \\
0.5$^*$ & 1.0\,$\pm$\,0.4 & 218\,$\pm$\,21 & 0.19 & 4.2\,$\pm$\,1.0 & 2.7\,$\pm$\,0.8 & 1.2\,$\pm$\,0.5 & 1.7\,$\pm$\,0.8 & 2.1\,$\pm$\,0.8 \\
0.6~ & 0.8\,$\pm$\,0.3 & 221\,$\pm$\,21 & 0.23 & 3.4\,$\pm$\,0.8 & 2.2\,$\pm$\,0.6 & 1.3\,$\pm$\,0.6 & 1.6\,$\pm$\,0.8 & 2.1\,$\pm$\,0.8 \\
0.7~ & 0.6\,$\pm$\,0.2 & 224\,$\pm$\,22 & 0.44 & 2.9\,$\pm$\,0.7 & 1.8\,$\pm$\,0.5 & 1.4\,$\pm$\,0.6 & 1.5\,$\pm$\,0.7 & 2.1\,$\pm$\,0.8 \\
0.8~ & 0.5\,$\pm$\,0.2 & 227\,$\pm$\,23 & 0.80 & 2.6\,$\pm$\,0.5 & 1.6\,$\pm$\,0.4 & 1.5\,$\pm$\,0.6 & 1.4\,$\pm$\,0.6 & 2.2\,$\pm$\,0.7 \\
\tableline 
\end{tabular}
\end{center}
\tablenotetext{1}{We assumed a flux calibration uncertainty of 20\% for each line. The uncertainty on the D$_2$O and H$_2^{18}$O column densities include both the flux calibration uncertainty and the excitation temperature uncertainty. For the determination of the ratios, we assumed ortho/para(D$_2$O) = 2, ortho/para(H$_2^{18}$O) = 3, and H$_2^{16}$O/H$_2^{18}$O = 560.}
\tablenotetext{*}{Best-fit obtained with the $\chi^2$ minimization.}
\end{table*}

\section{Discussion}

\subsection{High D$_2$O/HDO ratio vs low HDO/H$_2$O ratio}

Statistically, the replacement of an H atom by a D atom would lead to
a D$_2$O/HDO ratio 4 times lower than the HDO/H$_2$O ratio.
Surprisingly, we derived, in the inner region of NGC1333~IRAS2A, a
D$_2$O/HDO ratio a factor 7 higher than the HDO/H$_2$O ratio. Even
considering the  estimated uncertainties from the analysis (see
  Table \ref{table_RD}), a significant difference is observed between
these ratios.

A lower D$_2$O/HDO ratio could be obtained if the ortho/para ratio of
D$_2$O were below the LTE value of 2. Indeed, a value of
$\sim$1.1--1.3 was derived in the cold absorbing layer of
IRAS~16293-2422 \citep{Vastel2010,Coutens2013a}. For an ortho/para
ratio equal to 1, the D$_2$O/HDO ratio would be about 9\,$\times$\,10$^{-3}$. The
  mechanisms leading to an ortho/para ratio of D$_2$O lower than 2
  are, however, not understood. There is therefore no evidence
that the ortho/para ratio in the warm gas would be the same as in the
cold gas.

The HDO/H$_2$O ratio of $(1.7\pm 0.8)\times 10^{-3}$ derived here is a
factor of 2 higher than the LTE estimate without beam dilution by \citet{Persson2014} and a
factor of 2 lower than the non-LTE estimate by \citet{Taquet2013} for a
H$_2$ density of 10$^8$~cm$^{-3}$.
\citet{Persson2014} obtain a higher HDO/H$_2$O ratio of
$\sim$(2--3)\,$\times$\,10$^{-3}$ with a non-LTE excitation and
  radiative transfer model using a spherical envelope model with a
  temperature and density gradient constrained from other
  observations.
 Even
  considering HDO/H$_2$O ratios of
  $\sim$(2--3)\,$\times$\,10$^{-3}$ derived in these two studies, our
  observed D$_2$O/HDO ratio is still a factor 4--6 higher.
  
 The highest difference of a factor of 7 between the HDO/H$_2$O
  and D$_2$O/HDO ratios is found for the inner regions of
  NGC1333~IRAS2A.  In contrast, the D$_2$O/HDO ratio is only a factor
  of 2 higher than HDO/H$_2$O for the cold absorbing layer of
  IRAS16293-2422 \citep{Coutens2012,Coutens2013a}, while it is a
  factor of 2 lower for the hot core of Orion KL
 \citep[D$_2$O/HDO$\sim$1.6\,$\times$\,10$^{-3}$ vs.\
  HDO/H$_2$O$\sim$3\,$\times$\,10$^{-3}$;][]{Neill2013}.

Water observed in low-mass sources is formed mostly on the grains
  and then thermally desorbed in the warm inner regions of protostars ($T_{\rm d} \gtrsim 100$\,K).
  The gas-phase water D/H ratios then reflect the water deuterium
  fractionation in the icy grain mantles formed earlier in the
  molecular cloud and/or prestellar core by hydrogenation (and
  deuteration) of atomic or molecular oxygen. In the earliest
  molecular cloud phase, the deuterium fractionation is less than in the cold
  and dense prestellar stage, so the ice mantles are expected to have
  a layered structure, with the highest deuteration fractions on top
  \citep{Taquet2014}.  In the outer regions, these top ice layers
  are photodesorbed resulting in high gas-phase HDO and D$_2$O abundances.
  In the inner envelope, the entire ice mantle thermally desorbs
  and the gas phase abundances reflect the lower, less deuterated ice layers.

  Quantitative predictions of the D$_2$O/HDO and HDO/H$_2$O ratios
  have been obtained with pseudo-time dependent chemical models in
  which the physical conditions are constant
  \citep{Taquet2013b,Albertsson2013} and with 1D dynamical models in
  which a parcel of gas experiences different physical conditions with
  time
  \citep{Cazaux2011,Coutens2013a,Aikawa2012,Taquet2014,Wakelam2014}.
  In the pseudo-time dependent models, the water observed in the inner
  region comes from the sublimation of the ice mantles and can
  subsequently evolve chemically by gas-phase reactions, while in the
  dynamical models, the molecules remain in the warm gas for such a
  short time due to the rapid infall that no further chemistry can
  take place.  None of these models predict D$_2$O/HDO ratios higher
  than HDO/H$_2$O ratios, contrary to what is observed in
  NGC1333~IRAS2A.

This discrepancy between models and observations could mean that the
water deuteration on grain surfaces is far from understood and that
some deuteration pathways are incorrect or missing in the different
chemical models. A second explanation could be that,
  in addition to the thermal desorption, water is also formed at high
  temperature ($\gtrsim$230\,K) through the following reactions:
\begin{equation}
\rm O + H_2 \rightarrow OH + H, 
\end{equation}
\begin{equation}
\rm OH + H_2 \rightarrow H_2O + H. 
\end{equation} 
These gas-phase reactions do not play a major role in the pseudo-time dependent models by \citet{Taquet2013b} and \citet{Albertsson2013}, as the temperatures considered were lower than 150\,K.
The thermal desorption of grain mantles would then result in the
production of high D$_2$O/HDO ratios that reflect the water formation
at low temperatures on grain surfaces, while the gas-phase reactions at
high temperature would produce a lot of H$_2$O and little deuterated
water, resulting in a low HDO/H$_2$O ratio.  In this interpretation,
the D$_2$O/HDO ratio determined in NGC1333~IRAS2A ($\sim$1.2\%) would
reflect the ratio present in the ices. This value is in agreement with
the predictions of a chemical model including the multilayer ice mantle
approach and coupled with a 1D dynamical model, where
$\sim$(0.5--3)\,$\times$\,10$^{-2}$ is found in the inner regions of a
typical Class 0 protostar \citep[][]{Taquet2014}. It is also similar
to the upper limit constrained in the inner regions of IRAS~16293-2422
\citep[][$\le$0.9\%]{Coutens2013a}.

The high temperature gas-phase reactions (1) and (2) can take
  place both in the quiescent inner regions of protostellar envelopes
  where the temperature is above 230\,K and in shocks.  An extended
  component of H$_2^{18}$O attributed to outflows was detected towards
  NGC1333~IRAS2A \citep{Persson2012}. The base of the jet at the
  position of the protostar could possibly contribute to the
  H$_2^{18}$O flux. There is, however, no spectral evidence of
  that. The line profile of the H$_2^{18}$O line is narrow and similar
  to those of deuterated water. Besides, \citet{Persson2014} show
  that the HDO/H$_2$O ratios derived in NGC1333~IRAS2A are similar to
  NGC1333~IRAS4A and IRAS4B, but no extended emission of H$_2^{18}$O
  along outflows is detected towards these two sources. If these
  sources do not show outflow emission, their HDO/H$_2$O ratio should
  be higher, unless they present lower D$_2$O/HDO ratios than in
  NGC1333~IRAS2A.  H$_2$O could also be produced by accretion shocks
  when material falls onto the disk \citep{Neufeld1994},
  but as previously noted, the different water
  isotopologues show similar and relatively narrow line widths, which
  does not favor a shock scenario. Water formation in accretion shocks
  was also rejected for the low-mass protostar NGC1333~IRAS4B
  \citep{Jorgensen2010}. Consequently, if water is formed by gas-phase reactions at high temperature in low-mass protostars, it would probably take place in quiescent regions, rather than in
  shocks.
  
  These results highlight the importance of further experimental and theoretical studies of gas and grain surface deuteration. Dynamical effects
  can also be crucial for the chemistry. In 2D dynamical models including rotation and
  possibly disk formation, the time that the thermally desorbed
  molecules spend in the warm gas is lengthened considerably
  \citep[e.g.,][]{Visser2009} so that warm gas-phase chemistry becomes
  important in addition to the grain surface chemistry. 
  Such models should be attempted to check that the HDO/H$_2$O and D$_2$O/HDO ratios found in NGC1333~IRAS2A can be reproduced.
   Very high spatial resolution observations at 0.1$\arcsec$ scale may be able
to distinguish the various scenarios and fully rule out a shock
contribution. If H$_2^{18}$O shows enhanced emission in the innermost and
warmest regions (within $\sim$30 AU), while HDO and D$_2$O do not, it would
favor gas-phase formation of water at high temperatures. On the
contrary, if the different isotopologues show a similar distribution, it would imply a missing ingredient in our understanding of the surface deuteration process.

\subsection{The D$_2$O/HDO ratio from the cold outer to the warm inner regions}

No measurement of the D$_2$O/HDO ratio has been obtained in the
cold absorbing layer of NGC1333~IRAS2A. If we assume that it is
similar to the D$_2$O/HDO ratio measured in the cold layer surrounding
IRAS~16293-2422 ($\sim$11\%) or to the HDO/H$_2$O ratio derived
  in the cold envelope of NGC1333~IRAS2A by
  \citet[][$\sim$7\%]{Liu2011}, this would imply that the D$_2$O/HDO
  ratio decreases by a factor 6--10 from the cold to the warm
  regions.  This is consistent with the variations of the D$_2$O/HDO
ratio measured in IRAS~16293-2422 \citep{Coutens2013a}.

Such a decrease in the water deuterium fractionation from the cold to
the warm gas can be explained by the differential deuteration of
the multilayer ice mantles at the range of densities and temperatures
encountered prior to and during the star formation process
\citep{Taquet2014}.  If disks are present in the inner regions at
  the Class 0 stage, turbulent mixing could also explain some decrease
  of the water D/H ratio \citep{Furuya2013,Albertsson2014} but this
  would require a significant fraction of the emission to arise from
  the disk rather than envelope.

\acknowledgments

The authors thank the IRAM staff and especially Tessel van der Laan for their help with the observations and reduction of the data.

\end{document}